\documentclass[nofootinbib,aps,showpacs,twocolumn,amsmath,amssymb,unsortedaddress]{revtex4}
\usepackage{amssymb}
\usepackage{amsmath}
\usepackage{epsfig}
\usepackage{amsfonts}
\usepackage{color}
\usepackage{float}
\usepackage{latexsym}
\usepackage{mathrsfs}
\usepackage{feynmf}
\usepackage{balance}
\usepackage{multirow}
\usepackage{array}
\usepackage{multirow}
\usepackage[utf8]{inputenc}
\usepackage{longtable}

\usepackage{tikz}
\usetikzlibrary{shapes.geometric, arrows, positioning}

\begin{document}


\title{Three-Spin Systems and the Pusey-Barrett- Rudolph Theorem}
\author{Zeynab Faroughi}
\affiliation{Faculty of Physics, Shahid Bahonar University of Kerman, Kerman, Iran,}
\email{farougheez@sci.uk.ac.ir}
 \author{Ali Ahanj}
 \affiliation{Faculty of Science, Department of Physics, Khayyam University, Mashhad, Iran,}
  \email{a.ahanj@khayyam.ac.ir; ahanj@ipm.ir}
 \author{Samira Nazifkar}
\affiliation{Faculty of Science, Department of Physics, University of  Neyshabur, Neyshabur, Iran,}
\email{ nazifkar@neyshabur.ac.ir}
\author{Kurosh Javidan}
\affiliation{Faculty of Science, Department of Physics, Ferdowsi University of Mashhad, Iran}
\email{Javidan@um.ac.ir}
\date{\today}
\begin{abstract}
The fundamental nature of quantum wave function has been the topic of many discussions since the beginning of the quantum theory. It either corresponds to an element of reality $(\Psi-ontic)$ or  it is a subjective state of knowledge about the underlying reality $(\Psi-epistemic)$. Pusey, Barrett, and Rudolph (PBR) have shown  that epistemic interpretations of the quantum wave function are in contradiction with the predictions of quantum under some assumptions. In this paper, a laboratory protocol with a triple quantum dot will be introduced as a three-spin interaction system to study the PBR no-go theorem.  By this experimental model, we show that the epistemic interpretation of the quantum state is in contradiction with quantum theory, based only on the assumption that measurement settings can be prepared freely and independently from each other.
\end{abstract}


\maketitle
\section{\label{sec:level1}Introduction}
Since the beginning of quantum physics, one of the major problems  has been finding the relation between the quantum wave function  and the real physical world. The wave function (WF), as a mathematical description of the quantum state, carries all accessible knowledge about a quantum system. At first, the wave function was considered as a description of a real physical wave \cite{schr}. This idea has faced serious objections and has been rapidly replaced by Born's probability definition \cite{born}, which is the standard interpretation of the WF.  The standard interpretation does not tell us about the actuality of the physical world. Therefore, some alternative realistic interpretations of the WF have been presented and widely studied \cite{bohm1,bohm2, evertt,ghirardi,chiral} .
 Some investigations have proposed that the quantum state should be only a unique concept which contains the physical property of the quantum system \cite{rpm 1}. It may be noted that, in Bohmian interpretation of quantum mechanics, the WF is a part of reality along with hidden variables \cite{bohm1, bohm2}. A suitable framework in which these quantum states can be identified is the ontological models. In the ontological model of quantum theory, the well-defined set of the system of physical properties (ontic states)  is represented by a mathematical object, $\lambda$. The states $\lambda$ are called ontic states (or hidden variables). Furthermore, it is assumed that the pure quantum WF $|\psi\rangle$, induces a probability distribution $\mu_\psi(\lambda)$ on the ontic space of the system (denoted by $\Lambda$). The distributions $\mu_\psi(\lambda)$ are called epistemic states. In the realistic interpretations of quantum theory, the WF would be merely a description of the knowledge of the observer $(\Psi- epistemic)$, or it can be actually  interpreted as a state of reality that corresponds to the system $(\Psi-ontic)$.  Here, we will ignore the instrumentalist  interpretation of the quantum theory, in which the WF merely is a practical tool to predict credible solutions for a quantum system. What distinguished  $\Psi- epistemic$ and $\Psi-ontic$ ontological theories was formalized by Harrigan and Spekkans \cite{supcon}. According to their terminology, a model is called “ontic” if there does not exist any distinguished pair of quantum states which share the same physical properties $\lambda$. On the other hand, an epistemic model contains at least one ontic state which corresponds to more than one quantum state \cite{wave}. Now the important question that arises is  whether the quantum WF is an ontic object or an epistemic state. The answer to this question provides valuable insight for understanding the nature of the quantum state.\\
Pusey, Barret, and Rudolph (PBR) have attempted to get an answer to this fundamental question \cite{chiral}. They have introduced a completely novel ''no-go'' theorem which is formulated for an ontological model. They have shown that $\Psi-epistemic$ models cannot satisfy the predictions of quantum theory \cite{chiral, wave}. So the answer of the PBR to the above question, ruled out $\Psi-epistemic$ theorems and tried to provide a $\Psi-ontic$ view of the quantum state.

The key assumption of the PBR theorem is the preparation of independent postulation (PIP); however, the validity of this assumption is controversial \cite{rpm 1,bu, DUC, Farr, Wallden, Shane, patra, emerson}. There are some other propositions which contain weaker assumptions in comparison with basic PBR hypothesis \cite{bu}.  Colbeck and Renner got the same results of PBR but with a different argument \cite{rpm 1}. They have shown that the wave function of a quantum system is completely determined by all its elements of reality under the assumption of free-choice of measurement setting \cite{rpm 1}. Also Patra, Pironio and Massar \cite{patra} have argued that epistemic states are inconsistent with quantum predictions under continuity and a weak separability of remove of assumption. Hardy has presented another reasoning for the PBR theorem but with different assumptions\cite{hardy}. On the other hand it was shown theoretically by Barrett,  Cavalcanti, Lal, and Maroney \cite{maroney} and Leifer\cite{leifer} that the nonorthogonality of any two WF cannot be fully explained by a $\Psi-epistemic$ models for systems of dimension larger than 2. Next the authors in ref \cite{ringbaur} experimentally test this approach with single photons. In another work, Pati, Ghose and Rajagopal \cite{pati}  without using any additional assumption has shown that an epistemic WF that satisfies the Born rule is incompatible with the Schr\"{o}dinger time evolution. On the other hand, Lewis \emph{et al} \cite{lewis} have shown that under ignoring PIP and slightly weakens the definition of the epistemic state; it is possible to have an epistemic interpretation of quantum WF. Therefore, the situation is far from clear and continues to attract the physicists \cite{hardy, pati8,pati9,pati10,pati11,pati13, colbeck}.\\
Recently, the authors in \cite{bandyopadhyay} introduced a quantum exclusion game, first introduced in Ref. \cite{caves}, for analysing the PBR no-go theorem. The game of state exclusion has been explored under noisy channels \cite{heinosaari}, as well as its communication complexity properties \cite{perry, liu, ducuara}. Furthermore, the PBR theorem reformulated into a particular guessing game \cite{myrvold} and a Monty Hall game \cite{rajan}. However it seems that the feasible way to understand the validity of different interpretations of WF is experimental results based on possible laboratory proposals. To our best knowledge, there are only a few experimental proposal to test the PBR theorem using a special spin-spin interaction based on cold atom systems \cite{i22} and using trapped ions \cite{daniel}. Considerable attention has been paid to the experimental simulation of condensed matter systems, such as spin chains, because of rapid developments on optical lattice technologies. One can probe and realize complex quantum models with interesting properties in the laboratory by using such systems which can be realized by quantum dots (QD) \cite{qd}. New experimental activities are based on the three-spin interaction instead of older ising models. The three-spin model is widely used in recent activities in atomic and condensed matter physics \cite{QD, khaje, willke}. Thus, considering better models for many-body systems with three-spin interaction may help us to setup more practical experimental arrangements. Motivated by this situation, we have analyzed a triple QD system as a multi-component quantum system based on the three-spin interaction as a basic platform for experimental verification of epistemic/ontic interpretation of WF. Thus we present an effective Hamiltonian for the triple QD system containing external magnetic field-spin interaction as well as spin-spin and three-spins interactions. We will consider the state of the system as an experimental distribution over ontic state corresponding to the distinct pure and non-overlapping situation.\\
The article is organized as follows: In the Sec.II, the original proposals for the PBR theorem will be briefly reviewed. The three-spin interaction model is explained in the Sec.III, and finally our conclusions will be presented in Sec IV.
\section{\label{sec:level1} The original proposal for the PBR theorem }

 In this section, by applying the quantum state exclusion method \cite{perry,bandyopadhyay} the structure of the PBR no-go theorem will be reviewed as the first $\Psi-ontology$ theorem.
 To derive the theorem several assumptions are considered. The ontological assumption or realism is the first hypothesis in the PBR theorem. On the basis of this assumption, any underlying properties of the system (real or not real) are represented by a mathematical object $\lambda$ or, more generally its distribution $\mu (\lambda)$ \cite{chiral, wave}. Also under this hypothesis, as other no-go theorems like the Bell \cite{bell}, Kochen and Specker \cite{ks}, the real physical state is independent of the observer.  The second assumption is $\Psi-epistemism$,  which describes that the quantum wave function  can be only interpreted as our incomplete knowledge about the actual ontic state of the system.
The final required assumption to complete the proof of the PBR theorem is "preparation independence" (measurement independence).  It means that  when two independent systems are prepared separately, their physical states are also independent too. In other words, the properties described by  $\lambda$ are not correlated with the choice of measurement $M$.\
\begin{figure}
\centering
\includegraphics[scale=1]{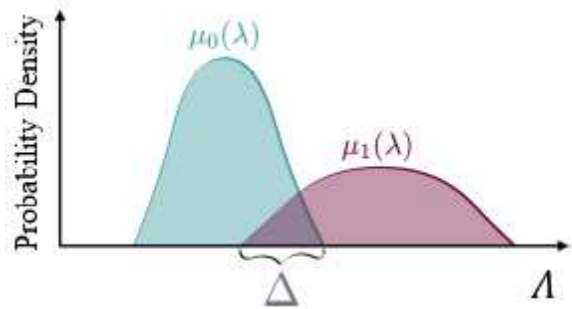}
\caption{\footnotesize
Two probability functions that represent quantum states with distributions that both assign positive probability
to some overlap region $\Delta$ }
 \label{sh1}
\end{figure}
In the simplified version of the PBR theorem, they  assume that the quantum state is indeed a state of knowledge and consider two distinct and non-orthogonal quantum states $|\Psi_0\rangle=|0\rangle$ and  $|\Psi_1\rangle=|+\rangle=1/\sqrt{2}(|0\rangle + |1\rangle) $  where $|0\rangle$ and $|1\rangle$ are orthonormal eigenfunctions of a two state system. The probability distributions $\mu_0(\lambda)$ and  $\mu_1(\lambda)$  related to $|\Psi_0\rangle$  and  $|\Psi_1\rangle$ have a common region $\Delta$  (see the Fig-1) according to the  $\Psi-epistemic$ model. It means that,  the overlap region $\Delta$ contains at least one common ontic state $\lambda_i \in \Delta$. Then there exists a positive parameter $q$ $(0<q<1)$ such that preparation of either quantum state $|\Psi_0\rangle$  and  $|\Psi_1\rangle$ results in a $\lambda_i \in \Delta$ with probability at least $q$  \cite{chiral}.
Now let's consider two separate systems with non-correlated physical states located in regions $A$ and $B$. Each of the systems $|\Psi_A\rangle$ and $|\Psi_B\rangle$ can be prepared independently in one of the states $|\Psi_0\rangle=|0\rangle$  and  $|\Psi_1\rangle=|+\rangle$ with equal probability. Thus, the four nonorthogonal  possible physical states of the two separately prepared systems are \cite{chiral}:
 \begin{equation}\label{1}
 |0\rangle_{A}|0\rangle_{B},~ |0\rangle_{A}|+\rangle_{B}, ~|+\rangle_{A}|0\rangle_{B} ,~  |+\rangle_{A}|+\rangle_{B}.
\end{equation}
Now, let's assume that the ontic states  $\lambda_A$ and $\lambda_B$ belong to the overlap region of the corresponding probability distributions $\mu_{0}(\lambda)$ and $\mu_{1}(\lambda)$ (i.e. $\lambda_A \in \Delta_A$ and $\lambda_B \in \Delta_B$ ) with probability $q^2$. This means that the physical state of the two systems  $A$ and $B$ is compatible with any of four possible quantum states which are given in (\ref{1}).
Let's  consider an entangled measurement acting on initially prepared systems which can be considered as a projection operator with following orthogonal states:
 \begin{eqnarray}\label{2}
|\xi_1\rangle=\frac{1}{\sqrt{2}}(|\;0\rangle_{A}|\;1\rangle_{B} + |\;1\rangle_{A} |\;0\rangle_{B})\nonumber\\
|\xi_2\rangle=\frac{1}{\sqrt{2}}(|\;0\rangle_{A}|-\rangle_{B} + |\;1\rangle _{A} |+\rangle_{B})\nonumber\\
|\xi_3\rangle=\frac{1}{\sqrt{2}}(|+\rangle_{A}|\; 1\rangle_{B} + |-\rangle_{A} |\; 0\rangle_{B})\nonumber\\
|\xi_4\rangle=\frac{1}{\sqrt{2}}(|+\rangle_{A}|-\rangle_{A}+|-\rangle_{A} |+\rangle_{B})
\end{eqnarray}
where $|-\rangle = \frac{1}{\sqrt{2}} (|0\rangle-|1\rangle)$.\\  By performing a set of measurements on prepared states (\ref{1}), the following results are achieved: $|\xi_1\rangle$ which is orthogonal to $|0\rangle_{A}|0\rangle_{B}$, $|\xi_2\rangle$ orthogonal to  $|0\rangle_{A}|+\rangle_{B}$, $|\xi_3\rangle$  orthogonal to  $|+\rangle_{A}|0\rangle_{B}$  and finally, $|\xi_4\rangle$ orthogonal to  $|+\rangle_{A}|+\rangle_{B} $. According to the quantum theory,  if initially prepared state is  $|0\rangle_{A}|0\rangle_{B}$,  then the result cannot be $|\xi_1\rangle$, i.e. the probability of finding $|\xi_1\rangle$ should be zero. Also, we can get the same result for $|\xi_2\rangle$, if the initial state is $|0\rangle_{A}|+\rangle_{B}$, $|\xi_3\rangle$ for the initial state $|+\rangle_{A}|0\rangle_{B}$ and finally $|\xi_4\rangle$, if the initial state is $|+\rangle_{A}|+\rangle_{B}$. No matter what result is obtained, it rules out one of the four preparations. But these results are a paradoxical outcome, because we assumed that all states should be found with an equal probability $q^2$.  It means that all final states can be produced with nonzero probability (and independent of initial prepared state) from $\lambda_i \in \Delta$, while the probability of finding $\lambda_i$ itself is $q^2$.
Through this reasoning, PBR have concluded that the states  $|0\rangle$ and  $|+\rangle$ cannot have any overlap region in the ontic state, so  the $\psi-epistemic$ model is not able to produce predictions of quantum theory.
In the next section, will be explained the three-spin interaction as a basic platform for an experimental proposal on the nature of WF.

\begin{figure}
\centering
\includegraphics[scale=1]{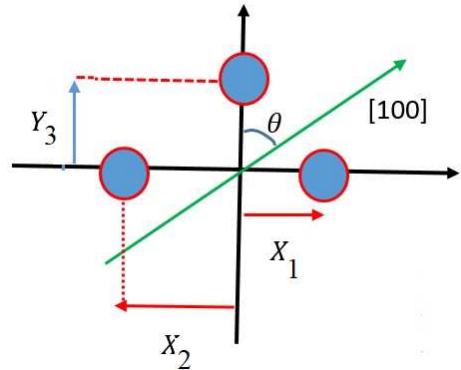}
\caption{\footnotesize
Triple QD in triangular configuration as three-spin interaction system. Distances of QDs respect to the center of mass are $X_{1}, X_{2}$ situated in the $X$ axis and the other $Y_{3}$ located in the direction of the $Y$ axis. The coordinate system has an angle $\theta$ with the $[100]$ direction of background crystalline environment.}
 \label{sh2}
\end{figure}
\section{\label{sec:level1} Triple quantum dot as a three-spin system}

In the following section, an experimental configuration will be introduced on a three-particle system as a proposal to test the PBR theorem. Such a set up has not been investigated so far. So it is better to start with explaining the general three-spin Hamiltonian sharing between three separated observers (for example, Alice, Bob, and Charlie).
A three-spin system is generally constructed by using an optical lattice combined with the technology of the cold atoms \cite{i23,Non-linear} and, also a triple quantum dot (QD) in linear or triangular configuration \cite{tqd,lqd}. Quantum dots as qubits are coupled with the effective spin Hamiltonian, which is defined by the geometrical arrangement of QDs, external fields and different interactions between system components. Geometrical behavior of QDs is defined by a potential which finds its minima at the position of dots. Spins interact through Coulomb force, spin-orbit interaction, and usually, move in an external magnetic field. Such a system is modeled through the orbital ground state of a two-dimensional harmonic oscillator in the external magnetic field.

Here we construct our model based on triple QD system as a three-spin configuration. Hamiltonian of the system depends on the spatial arrangement of QDs. There are two usual geometry of the triple dot in the laboratory usages: linear and triangular arrangements. In the triangular configuration, as shown in the Fig.2, two of QDs ( shown with 1 and 2 ) are positioned on the x-axis such that $x_1$ and $ x_2 $ are fixed, while the third QD (shown with 3) is located in the y-axis. As the QDs have a crystalline structure, one can arrange orientation of coordinate system with a certain angle with respect to the geometry of the crystalline axes of the substrate (for example with the 100 axes).
The effective spin Hamiltonian of three localized spin QD is written as \cite{QD, i23}:

\begin{eqnarray}\label{4}
H&=&\sum_{i=1,2,3} H_{i}^{(1)}(\mathbf{S_i})+ \sum_{i\neq j=1,2,3} H_{ij}^{(2)}(\mathbf{S_i},\mathbf{S_j})\nonumber\\ & +& H_{123}^{(3)}(\mathbf{S_1},\mathbf{S_2},\mathbf{S_3}).
\end{eqnarray}
 where $H^{(1)}$, $ H^{(2)}$ and $H^{(3)}$ denote single-, two- and three-spin interactions respectively. The single interaction appears with effective vector magnetic fields $\mathbf{b_i}$ as $ H_{i}^{(1)}=\mathbf{b_i}\cdot \mathbf{S_i}$ with $i=1,2,3 $ for QDs \cite{QD}. In practice, a single spin magnetic anisotropy locks each QD in a needed orientation which is tailored by the interaction with the crystal field \cite{khaje}. Thus, response of each QD to the magnetic field will be different, because of difference in g-factors ( QD magnetic moments ) and direction of spin orientations. Another method is adding a local magnetic field to each QDs \cite{willke}. The spin-spin interaction is divided into isotropic and anisotropic exchanges, but we consider only the isotropic part for simplicity as: $H_{i,j}^{(2)}=\mu_{ij}\mathbf{S_i}\cdot \mathbf{S_j}$ in which scalar parameters $\mu_{ij}$ are isotropic exchange couplings between QDs as $i\neq j=1,2,3 $. This means that isotropic spin-spin interaction is modeled as an effective term like: $H_{iso}=\mu_{12}\mathbf{S_1}\cdot \mathbf{S_2}+\mu_{13} \mathbf{S_1}\cdot \mathbf{S_3}+ \mu_{23} \mathbf{S_2}\cdot \mathbf{S_3} $. Three-spin interaction is modeled as: $H_{123}^{(3)}=\sum_{i,j,k}\gamma_{ijk}S_{1}^{i}S_{2}^{j}S_{3}^{k}$, where $i,j,k \in \{x,y,z\}$. The parameters  $\gamma_{ijk}$ are components of a direct product of three spin components, which are rank 3 tensors. As three spin system is an entangled three partite system, entanglement can flow within the system from one part to other parts. Thus, we do not need to setup an “at once” interaction between three QDs. It means that setting $\gamma_{ijk}$ to zero cannot decouple three QDs. Indeed we do not need non-zero parameter $\gamma_{ijk}$ necessarily since it can be explained by the general rotations of spins about each axis.
 We have chosen the following values for vectors of effective magnetic fields and isotropic couplings:
\begin{eqnarray}\label{z1}
\overrightarrow{b_1}= \left[ \begin {array}{c} -a\\ \noalign{\medskip}0\\ \noalign{\medskip}b\end {array} \right],\overrightarrow{b_2}=\left[ \begin {array}{c} c\\ \noalign{\medskip}0\\ \noalign{\medskip}0\end {array} \right],\overrightarrow{b_3}=\left[ \begin {array}{c} a\\ \noalign{\medskip}0 \\ \noalign{\medskip}b\end {array} \right],
\end{eqnarray}

and:
\begin{eqnarray}\label{z2}
\nonumber \mu_{12}&=&\left[ \begin {array}{ccc}\; a&\;0\;&a\\ \noalign{\medskip}0&a&0
\\ \noalign{\medskip}-a&0&c\end {array} \right], \mu_{13}=\left[ \begin {array}{ccc} 0&\;0\;&a\\ \noalign{\medskip}0&0&0
\\ \noalign{\medskip}-a&0&0\end {array} \right],\\
\mu_{23}&=&\left[ \begin {array}{ccc} a&0&-a\\ \noalign{\medskip}0&a&0
\\ \noalign{\medskip}-a&0&-c\end {array} \right].
\end{eqnarray}

in which $a, b$ and $c$ are real dimensionless free parameters are related to the appropriate units for magnetic field components, location, spin configurations, energy and other observables of QDs. Three spin interactions have been arranged by the rotation of $H_{23}$,$H_{13}$ and $H_{12}$  around $y$, $z$ and $x$ axes respectively. By using the above selections, the general form of Hamiltonian $H$ becomes ($\frac{\hbar}{2}=1$):

\begin{widetext}
 \begin{eqnarray}\label{f5}
H=2\left[ \begin {array}{cccccccc} \textrm {b}&\textrm {a}&\textrm {c}-\textrm {a}&-\textrm {a}&-\textrm {a}&0&-\textrm {a}&0\\ \noalign{\medskip}\textrm {a}&\textrm {c}&\textrm {a}&0&0&\textrm {a}&\textrm {a}-\textrm {b}&\textrm {a}
\\ \noalign{\medskip}\textrm {c}-\textrm {a}&\textrm {a}&\textrm {b}&\textrm {a}&\textrm {a}&0&-\textrm {a}&0\\\noalign{\medskip}-\textrm {a}&0&\textrm {a}&-\textrm {c}&\textrm {a}+\textrm {b}&\textrm {a}&0&-\textrm {a}\\ \noalign{\medskip}-\textrm {a}&0&\textrm {a}&\textrm {a}+\textrm {b}&-\textrm {c}&-\textrm {a}&0&\textrm {a}\\ \noalign{\medskip}0&\textrm {a}&0&\textrm {a}&-\textrm {a}&-\textrm {b}&\textrm {a}&\textrm {c}+\textrm {a}\\ \noalign{\medskip}-\textrm {a}&\textrm {a}-\textrm {b}&-\textrm {a}&0&0&\textrm {a}&\textrm {c}&\textrm {a}\\ \noalign{\medskip}0&\textrm {a}&0&-\textrm {a}&\textrm {a}&\textrm {a}+\textrm {c}&\textrm {a}&-\textrm {b}   \end {array} \right]
\end{eqnarray}\\

 The eigenvalues $E_i $ and eigenstates $|e_i\rangle$ of Hamiltonian  are  as follows:

\begin{eqnarray}\label{z9}
&&E_1= 2 \textrm {b}-2\textrm {c}-2\textrm {a},\quad\quad\quad|e_1\rangle=1/2(|010\rangle-|000\rangle-|011\rangle-|100\rangle)\nonumber\\
&&E_2=2\textrm {b}-2\textrm {c}+6\textrm {a},\quad\quad\quad|e_2\rangle=1/2(|000\rangle-|010\rangle-|011\rangle-|100\rangle)\nonumber\\
&&E_3=2\textrm {a}-2\textrm {b}-2\textrm {c},\quad\quad\quad|e_3\rangle=1/2(|011\rangle-|100\rangle+|101\rangle-|111\rangle)\nonumber\\
&&E_4= 2\textrm {c}-2\textrm {b}-6\textrm {a},\quad\quad\quad|e_4\rangle=1/2(|100\rangle-|011\rangle+|101\rangle-|111\rangle)\nonumber\\
&&E_5= 2\textrm {a}+2\textrm {b}+2\textrm {c},\quad\quad\quad|e_5\rangle=1/2(|000\rangle+|001\rangle+|010\rangle-|110\rangle)\nonumber\\
&&E_6=2\textrm {b}-6\textrm {a}+2\textrm {c},\quad\quad\quad|e_6\rangle=1/2(|000\rangle-|001\rangle+|010\rangle+|110\rangle)\nonumber\\
&&E_7= 2\textrm {c}-2\textrm {a}-2\textrm {b},\quad\quad\quad|e_7\rangle=1/2(|101\rangle-|001\rangle-|110\rangle+|111\rangle)\nonumber\\
&&E_8= 2\textrm {c}+6\textrm {a}-2\textrm {b},\quad\quad\quad|e_8\rangle=1/2(|001\rangle+|101\rangle+|110\rangle+|111\rangle)
\end{eqnarray}\\
\end{widetext}

It should be noticed that, by choosing $\mid a\mid \neq \mid b\mid \neq \mid c \mid$, the degenerate states will be resolved. Laboratory prepration of initial condition for spin orientation in QDs is a well known process \cite{A, B, C,D}.

Now, let's take two arbitrary distinct nonorthogonal quantum states $| m \rangle$ and $| n \rangle$ as follows:
\begin{eqnarray}\label{z6}
|m\rangle = \cos(\theta/2)|0\rangle - \sin(\theta/2)|1\rangle\nonumber\\
|\;n \rangle = \cos(\theta/2)|0\rangle + \sin(\theta/2)|1\rangle
\end{eqnarray}
where $\mid \langle m| n \rangle \mid ^2 = \cos ^2\theta $ $( 0< \theta < \frac{\pi}{2} )$ and the quantum states orthogonal to above states are:
\begin{eqnarray}\label{z7}
|\overline{m}\rangle=\; \;\:  \sin(\theta/2)|0\rangle + \cos(\theta/2)|1\rangle\nonumber\\
|\overline{n}\rangle= -\sin(\theta/2)|0\rangle + \cos(\theta/2)|1\rangle.
\end{eqnarray}
With every run of the experiment, Alice produces one of the states $| m\rangle$ or $| n\rangle$; Bob prepares one of the states $|m\rangle$ or $|\overline{n}\rangle$;  and also Charlie produces one of the states $|\overline{m}\rangle$ or $|\overline{n}\rangle$ independently. Since these people are preparing their states in their space-like separated regions, there is not any operational correlation between their preparation procedures and then $\psi_{ABC}=\psi_A \otimes \psi_B \otimes \psi_C$.
So, the eight possibilities for the preparation of the composite system $\psi_{ABC}$ between Alice, Bob and Charlie are as follows:
\begin{eqnarray}\label{z8}
\Psi^{(1)}_{ABC}&=&|m\rangle_A |m\rangle_B\:  |\overline{m}\rangle_C  \nonumber\\
\Psi^{(2)}_{ABC}&=&|m\rangle_A |m\rangle_B\;  |\overline{n}\; \rangle_C\nonumber\\
\Psi^{(3)}_{ABC}&=&|m\rangle_A  | \overline{n}\rangle_B\; \,  \, |\overline{m}\rangle_C     \nonumber\\
\Psi^{(4)}_{ABC}&=&|m\rangle_A  | \overline{n}\rangle_B\;\,   \, | \overline{n}\; \rangle_C\nonumber\\
\Psi^{(5)}_{ABC}&=&|n \rangle_A \: |m\rangle_B\:  \: |\overline{m}\, \rangle_C        \nonumber\\
\Psi^{(6)}_{ABC}&=&|n \rangle_A\:  |m\rangle_B\:  \:|\overline{n}\; \rangle_C\nonumber\\
\Psi^{(7)}_{ABC}&=&|n \rangle_A\: |\overline{n}\rangle_B\; \, \; |\overline{m}\,\rangle_C     \nonumber\\
\Psi^{(8)}_{ABC}&=&|n \rangle_A\:  |\overline{n}\rangle_B\; \,  \; |\overline{n}\;   \rangle_C.
\end{eqnarray}
Now these three prepared atoms are interacted through the Hamiltonian (\ref{f5}).
\begin{figure}[htp]
\centering
\includegraphics[scale=0.8]{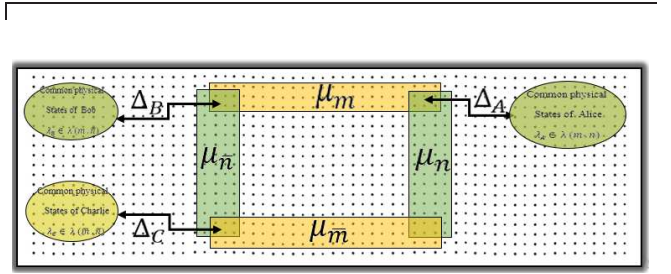}
\caption{\footnotesize
 It is assume that $|m\rangle$ and $|n\rangle$, $|m\rangle$ and $|\overline{n}\rangle$, and $|\overline{m}\rangle$ and $|\overline{n}\rangle$ are conjoint. The joint physical state Alice,Bob and Charlie :               $\lambda_{ABC}=\lambda_A \lambda_B \lambda_C$.  Let's consider a run of the experiment in which  $\lambda_A\in\lambda(m,n)$, $\lambda_B\in\lambda(m ,\overline{n})$ , and $\lambda_C\in\lambda(\overline{m}, \overline{n })$  }
 \label{sh3}
\end{figure}
The eigenstates $|e_i\rangle$ of the Hamiltonian (\ref{f5}) are non-degenerate  Bell states with different eigenvalues $E_i$ which are given in (\ref{z9}).
Now, we define the measurement operator $\hat{M}^{(i)}_{ABC}\equiv |e^{i}\rangle \langle e^{i}| $ with eight possible outcomes $m=1,\ldots 8$. Also, each of the orthonormal measurement bases $\{ |e_{i}\rangle\}$ is orthogonal to one of the eight preparation states $\psi ^{(i)}_{ABC}$ ( $\langle e_i | \psi^{(i)}_{ABC} \rangle = 0$ ).
 So, according to the Born's rule in standard quantum mechanics, the conditional probability of finding the $\hat{M} ^{(i)}$ measurement outcome on the state $\rho _i \equiv |\psi^{(i)}_{ABC}\rangle \langle \psi^{(i)}_{ABC} |$
 should be zero  written as  follows:
\begin{eqnarray}\label{f1}
pr(m^{ABC}=1|\Psi^{(1)}_{ABC},\quad \hat{M}^{(1)}_{ABC})=0\nonumber\\
pr(m^{ABC}=2|\Psi^{(2)}_{ABC},\quad \hat{M}^{(2)}_{ABC})=0\nonumber\\
pr(m^{ABC}=3|\Psi^{(3)}_{ABC},\quad \hat{M}^{(3)}_{ABC})=0\nonumber\\
pr(m^{ABC}=4|\Psi^{(4)}_{ABC},\quad \hat{M}^{(4)}_{ABC})=0\nonumber\\
pr(m^{ABC}=5|\Psi^{(5)}_{ABC},\quad \hat{M}^{(5)}_{ABC})=0\nonumber\\
pr(m^{ABC}=6|\Psi^{(6)}_{ABC},\quad \hat{M}^{(6)}_{ABC})=0\nonumber\\
pr(m^{ABC}=7|\Psi^{(7)}_{ABC},\quad \hat{M}^{(7)}_{ABC})=0\nonumber\\
pr(m^{ABC}=8|\Psi^{(8)}_{ABC},\quad \hat{M}^{(8)}_{ABC})=0\nonumber\\
\end{eqnarray}

The ontological model of the quantum theory assumes that the ontic state of the composite system  $\psi_{ABC}=\psi_A  \psi_B \psi_C$ can be written in the following way:
$$\Lambda_{ABC}=\Lambda_A \times \Lambda_B \times \Lambda_C$$
where $\lambda_{ABC} \in \Lambda_{ABC}$ is the ontic state of the tripartite system $ABC$. Now, let's consider the quantum states $|m\rangle$ and $|n\rangle$ which contain sets of  the physical states $\lambda_m$ and $\lambda_n$ respectively. $\lambda_m$ and $\lambda_n$ determine the outcomes of experiments performed on the quantum states. We denote all of common physical states of $|m\rangle$ and $|n\rangle$ by $\lambda(m,n)$. It is important to notice that two orthogonal quantum states $|m\rangle$ and $|n\rangle$ cannot share the same physical states, i.e. $\lambda(m,n)$ is empty. In the $\Psi-epistemic$ interpretation, the overlap region between the probability distributions related to non-orthogonal (distinct) states $|m\rangle$ and $|n\rangle$ contains at least one common ontic state $\lambda_i \in \lambda(m,n)$.

According to the Fig.3, the quantum states $|m\rangle$ and $|n\rangle$, $|m\rangle$ and $|\bar{n}\rangle$, $|\bar{m}\rangle$ and $|n\rangle$, $|\bar{m}\rangle$ and $|\bar{n}\rangle$ have overlaps $\lambda(m,n)$, $\lambda(m,\bar{n})$, $\lambda(\bar{m},n)$ and  $\lambda(\bar{m},\bar{n})$ in the probability distributions over their ontic state spaces respectively. Now let's consider a particular region of the ontic state space such that $\lambda_A \in \lambda(m,n)$, $\lambda_B \in \lambda(m,\bar{n})$, and $\lambda_C \in \lambda(\bar{m},\bar{n})$. We notice that, the mentioned particular region of $\Lambda_{ABC}$ produces all states (\ref{z9}) with a nonzero probability indeed. In other words, we cannot distinguish  which of eight states is responsible for what the measuring device creates and, therefore, this is a problematic region in $\psi-epistemic$ models. Then our experimental proposal based on the three-spin Hamiltonian is able to produce the results of the PBR theorem.

\section{\label{sec:level1} Conclusion}
A brief review of the original proposal of the Pusey, Barret and Rudolph and the contradiction between $\psi-epistemic$ approach and the quantum mechanics has been presented.  An alternative experimental proposal based on the three-spin interaction Hamiltonian has been suggested. We have proposed a triple QD configuration as a three-spin interaction which is widely used in new experimental setups. Thus, such systems may help us to design a tripartite quantum system for investigating the PBR theorem. Suitable Hamiltonian feasible for examining the PBR theorem is constructed, by calculating particular values for the system parameter like an applied magnetic field on atoms, the geometrical configuration of cold atoms and spin-spin interactions. In addition to the benefits of possible experimental applications, presented model, itself is also worthwhile as a theoretical platform for examining the logical aspects of quantum mechanics. Thus we have presented a theoretical evaluation for such a possible experimental setup and discussed it's results based on $\psi$-epistemic and $\psi$-ontic.  It seems impossible to reproduce all the results of quantum theory for the presented setup by considering $\psi-epistemic$ approach; the mentioned contradiction reveals that $\psi$ should not be interpreted merely as an epistemic object.

\section*{\label{sec:level1} Acknowledgement}
The authors would like to thank professor Omid Ghahreman and Ali Pourvali for their comments that
greatly improved the manuscript.

\providecommand{\noopsort}[1]{}\providecommand{\singleletter}[1]{#1}%

\end{document}